\documentclass[superscriptaddress]{revtex4-2}
\usepackage{amssymb,amsmath,epsfig}
\usepackage[colorlinks=true, pdfstartview=FitV, linkcolor=blue, citecolor=red, urlcolor=magenta, breaklinks=true]{hyperref}
\usepackage{subfigure}
\usepackage{amsfonts}
\usepackage{graphicx,epsfig} 
\usepackage{subfigure}

\begin{document}
\title{The self-dual Lorentz violating model: quantization, scattering and dual equivalence}

\author{M. A. Anacleto}\email{anacleto@df.ufcg.edu.br}
\affiliation{Departamento de F\'{\i}sica, Universidade Federal de Campina Grande
Caixa Postal 10071, 58429-900 Campina Grande, Para\'{\i}ba, Brazil}
\author{F. A. Brito}
\email{fabrito@df.ufcg.edu.br}
\affiliation{Departamento de F\'{\i}sica, Universidade Federal de Campina Grande
Caixa Postal 10071, 58429-900 Campina Grande, Para\'{\i}ba, Brazil}
\affiliation{Departamento de F\'isica, Universidade Federal da Para\'iba, 
Caixa Postal 5008, 58051-970 Jo\~ao Pessoa, Para\'iba, Brazil}
\author{E. Passos}
\email{passos@df.ufcg.edu.br}
\affiliation{Departamento de F\'{\i}sica, Universidade Federal de Campina Grande
Caixa Postal 10071, 58429-900 Campina Grande, Para\'{\i}ba, Brazil}

\begin{abstract}
In this paper, we analysis the dynamics, at the quantum level, of the self-dual field minimally coupled to bosons with Lorentz symmetry breaking. We quantize the model by applying the Dirac bracket canonical quantization procedure. In addition, we test the relativistic invariance of the model by computing the boson-boson elastic scattering amplitude. 
Therefore, we show that the Lorentz symmetry breaking has been restored at the quantum level. 
We finalize our analysis by computing the dual equivalence between the self-dual model with Lorentz symmetry breaking coupled with bosonic matter and the Maxwell-Chern-Simons with Lorentz invariance violation coupled with bosonic field.
\end{abstract}
\maketitle
\pretolerance10000

\section{Introduction}
In physics, symmetries play important roles. 
In particular, the violation of Lorentz symmetry initiated by Kostelecky and collaborators~\cite{Kostelecky:1988zi} has been widely investigated in various fields.
The motivation comes from the theory of superstrings indicating that the Lorentz symmetry would be broken at high energies.
It is well known that Lorentz symmetry is of fundamental importance in the construction of physical theories. Therefore, this symmetry was tested with great precision~\cite{Kostelecky:2008ts}. 
From a theoretical point of view, it is conjectured that violations of Lorentz symmetry can occur on the Planck scale. 
In addition, these effects are observed even in the low-energy region. 
In~\cite{Carroll:1989vb}, the breaking of Lorentz symmetry in field theory was studied. 
Lorentz symmetry breaking in QED was investigated in~\cite{Kostelecky:2001jc,Gomes:2007rv,Nascimento:2007rb}. 
Here, our focus of attention is on the self-dual field model coupled with bosons, in the background violating Lorentz symmetry.

The self-dual field model proposed by Townsend, Pilch and Van Nieuwenhuizen~\cite{Townsend:1983xs} has received a lot of attention.
The equivalence between the free self-dual model and the Maxwell Chern-Simons model was established in~\cite{Deser:1984kw,Townsend:1984}. 
Also, in~\cite{Banerjee:1995yf}, equivalence at the level of the Green function has been established. 
In~\cite{Gomes:1997mf}, Gomes and collaborators explore the equivalence between the self-dual model minimally coupled with a fermion field and the Maxwell Chern-Simons model coupled with fermions.  
In addition, the quantization of the self-dual model coupled with fermions and the relativistic invariance test via calculation of the fermion-fermion scattering amplitude was performed~\cite{Girotti:1998hb}. 
Furthermore, the quantization of the self-dual model coupled with bosons, as well as the relativistic invariance test computing the boson-boson scattering was also studied in~\cite{Anacleto:2000ea}. 
Later, the dual equivalence of the self-dual model coupled with bosons and the Maxwell Chern-Simons model coupled with bosons was established~\cite{Anacleto:2001rp}. 
An analysis on the coupling of the self-dual field to dynamic U(1) matter and its dual theory has been carried out in~\cite{Dalmazi:2003sz} 
Duality in the context of Lorentz symmetry breaking has been investigated in four-dimensional~\cite{BottaCantcheff:2003wh,Guimaraes:2006gj,Guimaraes:2010cu} and three-dimensional theories~\cite{Furtado:2008gs}. 
Besides, different aspects of duality were constructed~\cite{Ferrari:2006vy,Karlhede:1986qd,Bralic:1995ip,Banerjee:1996nq,Ilha:2001he}. 
Also see~\cite{Banerjee:2000hs,Menezes:2003vz,Bazeia:2002bz,Mariz:2003vx,Dalmazi:2005bp,Guimaraes:2005hd,Dalmazi:2006yv,Dalmazi:2006bc} for other applications.
Recently, in~\cite{Belchior:2023cqt} the duality between the Maxwell-Chern-Simons and self-dual models in very special relativity has been investigated.

In this paper, we are interested in studying the canonical quantization of the self-dual field model coupled with bosons violating Lorentz symmetry.
In this study, we will quantize the model by applying the canonical quantization procedure of the Dirac brackets. 
Besides, we will compute the boson-boson scattering amplitude to test the Lorentz invariance of the model. 
Therefore, as a result of this analysis, we show that the combined action of the non-covariant parts of the interaction Hamiltonian is replaced by the minimum covariant field-current interaction. 
In addition, we verify that Lorentz invariance is preserved at the quantum level. 
Moreover, we investigate the duality between the self-dual and Maxwell-Chern-Simons models with Lorentz symmetry breaking and coupled with bosonic matter.

The paper is organized as follows. In Sec. \ref{sdmvl} we introduce the Lorentz symmetry breaking effect in the self-dual field model coupled with bosons and quantize the model via Dirac brackets. In Sec. \ref{scat} we compute the boson-boson elastic scattering amplitude to test the Lorentz invariance of the model. 
In Sec. \ref{Dual} we explore the dual equivalence between the self-dual and Maxwell-Chern-Simons models with Lorentz symmetry breaking and coupled with bosonic matter.
Finally in Sec.~\ref{summ} we present our final considerations.

\section{The Lorentz-breaking self-dual model}
\label{sdmvl}
In this section, we consider the self-dual field model coupled to bosons with Lorentz symmetry breaking, given by the following Lagrangian:
\begin{eqnarray}
\label{acao}
\mathcal{L}=\mathcal{L}_{sd} + \mathcal{L}_{\phi f} + \mathcal{L}_{\varphi}  + \mathcal{L}_{int},
\end{eqnarray}
being
\begin{eqnarray}
\mathcal{L}_{sd}&=&\frac{m}{2} \epsilon^{\mu\nu\rho}(f_{\mu}\partial_{\nu}f_{\rho})-\frac{m^2}{2}f_{\mu}f^{\mu},
\\
\mathcal{L}_{\phi f}&=&\frac{1}{2}\partial_{\mu}\phi\partial^{\mu}\phi+ 2m\phi v^{\mu}f_{\mu},
\\
\mathcal{L}_{\varphi}&=&\partial_{\mu}\varphi^{\ast}\partial^{\mu}\varphi-M^2\varphi^{\ast}\varphi,
\\
\mathcal{L}_{int}&=&-ef_{\mu}J^{\mu}+ e^2f_{\mu}f^{\mu}\varphi^{\ast}\varphi-e^2\phi^2\varphi^{\ast}\varphi,
\end{eqnarray}
where $J_{\mu}=i(\varphi^{\ast}\partial_{\mu}\varphi-\partial_{\mu}\varphi^{\ast}\varphi)$ is the current, $ m $ is a parameter with dimensions of mass, $ \varphi $ is the charged scalar field, $M$ is the boson mass, $ f_{\mu} $ is the self-dual field and $ \phi $ is the real scalar field interacting with the self-dual field $f_{\mu}$. 
Here, $2m\phi v^{\mu}f^{\mu}$ is the term carrying the Lorentz symmetry breaking with the constant 3-vector introducing a preferred frame of reference in spacetime. 
The model consisting of $\mathcal{L}_{sd}$ and $\mathcal{L}_{\phi f}$ was considered in~\cite{Furtado:2008gs} to examine the self-dual/Maxwell-Chern-Simons duality with Lorentz symmetry-breaking. 
{However, we emphasize that a study of the self-dual model with Lorentz symmetry breaking coupled with bosonic matter has not been performed. Hence, we will consider the contribution of the boson field to the Lagrangian density ($\mathcal{L}_{\varphi}$ and $\mathcal{L}_{int}$) to explore quantization, scattering and dual equivalence.}
In the Maxwell-Chern-Simons model with Lorentz symmetry breaking, the mixed term, $\phi\epsilon^{\mu\nu\rho}F_{\nu\rho}$, was obtained through the dimensional reduction of the {Carroll-Field-Jackiw (CFJ) term}~\cite{Belich:2002vd,Belich:2004pi,Belich:2003xa}. 
For other applications of this term, see also~\cite{Gaete:2005ba,Gaete:2006ss,Anacleto:2015nra}.
{In addition, Casana and collaborators successfully employed the Gupta-Bleuler quantization technique with Lorentz gauge corrected with the contribution of Lorentz symmetry violation in modified Maxwell electrodynamics~\cite{Casana:2014cqa,Casana:2016sfy}.}

Now we start the development of the canonical quantization approach of the self-dual model coupled to bosons and with Lorentz symmetry breaking.  
Then, by computing the canonically conjugate moments of the model, we find
\begin{eqnarray}
\pi_{\alpha}&=&\frac{\partial{\cal L}}{\partial(\partial_{0}f_{\alpha})}=-\frac{m}{2}\epsilon_{0\alpha\mu}f^{\mu},
\\
\Pi_{\phi}&=&\frac{\partial{\cal L}}{\partial(\partial_{0}\phi)}=\partial_{0}\phi,
\\
{\Pi}&=&\frac{\partial{\cal L}}{\partial(\partial_{0}\varphi)}=\partial_{0}\varphi^{\ast}-ief^{0}\varphi^{\ast},
\\
{\Pi}^{\ast}&=&\frac{\partial{\cal L}}{\partial(\partial_{0}\varphi^{\ast})}=\partial_{0}\varphi + ief^{0}\varphi.
\end{eqnarray}
Hence, we find that the primary constraints are given by
\begin{eqnarray}
\Phi^{(1)}_{0}&=&\pi_{0}\approx 0,
\\
\Phi^{(1)}_{i}&=&\pi_{i}+\frac{m}{2}\epsilon_{ij}f^{j}\approx 0.
\end{eqnarray}
According to the Dirac quantization procedure, the symbol ($\approx$) introduced above means weak equality and with equality occurring on the constraint surface~\cite{Dirac}. 
However, the canonical Hamiltonian density is given by
\begin{eqnarray}
\label{ham}
{\cal H}&=&\Pi_{\phi}\partial_{0}\phi+\Pi\partial_{0}\varphi+\Pi^{\ast}\partial_{0}\varphi^{\ast}
+\pi_{i}\partial_{0}f^{i}-{\cal L},
\nonumber\\
&=&\Pi^{\ast}\Pi + \frac{\Pi_{\phi}^2}{2}-\frac{1}{2}\partial_{i}\phi\partial^{i}\phi
-\partial_{i}\varphi^{\ast}\partial^{i}\varphi+M^2\varphi^{\ast}\varphi+e^2\phi^2\varphi^{\ast}\varphi
+ef^{i}J_{i}-e^2f_{i}f^{i}\varphi^{\ast}\varphi
\nonumber\\
&+&\frac{m^2}{2}f_{\mu}f^{\mu}-2m\phi{v_{i}f^{i}}
-f^{0}[m\epsilon_{ij}\partial^{i}f^{j}+ ie(\Pi\varphi-\Pi^{\ast}\varphi^{\ast})+2m\phi{v_{0}}].
\end{eqnarray}
Thus, the primary Hamiltonian can be presented as
\begin{eqnarray}
H_{p}=\int d^2x({\cal H}+U^{0}\Phi^{(1)}_{0}+U^i\Phi^{(1)}_{i}),
\end{eqnarray}
being $U^{0}$ and $U^{i}$ the Lagrange multipliers.

Next, we apply the Dirac algorithm by imposing the condition of conservation in time, such that:
\begin{eqnarray}
\dot{\Phi}^{(1)}_{0}=\{\Phi^{(1)}_{0},H_{p}\}_{p}=\{\pi_{0}(\vec{x}),H_{p}(\vec{y})\}_{p}\approx 0.
\end{eqnarray} 
This condition gives rise to the secondary constraint
\begin{eqnarray}
\label{vincsec}
\Phi^{(2)}&=&m^2\left(f^{0}-\frac{1}{m}\epsilon_{ij}\partial^{i}f^{j}
-\frac{ie}{m^2}(\Pi\varphi-\Pi^{\ast}\varphi^{\ast})-\frac{2\phi{v^0}}{m}\right)\approx 0.
\end{eqnarray}
Then, to complete the Dirac algorithm we must again check whether the conservation of the constraints $\Phi^{(2)}$ implies new secondary constraints. It is easy to see that no more constraints arise and the Lagrange multipliers are determined. We can verify that the system of constraints is of the second class. 

At this point, we begin the process of quantizing the theory with second-class constraints by introducing the Dirac brackets defined for arbitrary functions of fields and momenta as follows:
\begin{eqnarray}
\{\Lambda,\Omega\}_{D}=\{\Lambda,\Omega\}
-\sum_{q=0}^{3}\sum_{p=0}^{3}\int\int{d^2ud^2v\{\Lambda,\xi_{p}(\vec{u})\}}R^{-1}_{pq}\{\xi_{q}(\vec{v}),\Omega\},
\end{eqnarray}
where $\xi_{0}\equiv \Phi^{(1)}_{0}$, $\xi_{l}\equiv \Phi^{(1)}_{l}, l=1,2$ , $\xi_{3}\equiv \Phi^{(2)}$ and $R_{pq}(\vec{u},\vec{v})=\{\xi_{p}(\vec{u}),\xi_{q}(\vec{v})\}$ is a matrix of the Poisson brackets of the constraints (it is non-degenerate as it should be for the system of the second-class constraints).

In the sequence to quantize the theory, we define the fields and momenta as being operators and thus we have the following relation involving the commutator and Dirac bracket:
\begin{eqnarray}
[\Lambda,\Omega]=i\{\Lambda,\Omega\}_{D}.
\end{eqnarray}
Thus, after computing the Dirac brackets, we find the following commutation relations at equal times for the dynamic variables
\begin{eqnarray}
&&[f^{0}(\vec{x}),f^{j}(\vec{y})]=i\partial^{j}_{x}\delta(\vec{x}-\vec{y}),
\nonumber\\
&&[f^{k}(\vec{x}),f^{j}(\vec{y})]=-im\epsilon^{kj}\delta(\vec{x}-\vec{y}),
\nonumber\\
&&[f^{0}(\vec{x}),\pi_{k}(\vec{y})]=-\frac{i}{2m}\epsilon_{kj}\partial^{j}_{x}\delta(\vec{x}-\vec{y}),
\nonumber\\
&&[f^{j}(\vec{x}),\pi_{k}(\vec{y})]=\frac{i}{2}g_{k}^{j}\delta(\vec{x}-\vec{y}),
\nonumber\\
&&[\pi_{j}(\vec{x}),\pi_{k}(\vec{y})]=-\frac{i}{4m}\epsilon_{jk}\delta(\vec{x}-\vec{y}),
\nonumber\\
&&[f^{0}(\vec{x}),\varphi(\vec{y})]=e\varphi\delta(\vec{x}-\vec{y}),
\nonumber\\
&&[f^{0}(\vec{x}),\varphi^{\dagger}(\vec{y})]=-e\varphi^{\dagger}\delta(\vec{x}-\vec{y}),
\nonumber\\
&&[\phi(\vec{x}),\Pi_{\phi}(\vec{y})]=i\delta(\vec{x}-\vec{y}),
\nonumber\\
&&[\varphi(\vec{x}),\Pi(\vec{y})]=i\delta(\vec{x}-\vec{y}),
\nonumber\\
&&[\phi^{\dagger}(\vec{x}),\Pi^{\dagger}(\vec{y})]=i\delta(\vec{x}-\vec{y}).
\end{eqnarray}

The Hamiltonian operator (\ref{ham}), with use of the constraint (\ref{vincsec}), can be rewritten as
\begin{eqnarray}
H&=&\int{d^2x}\left(\Pi^{\dagger}\Pi + \frac{\Pi_{\phi}^2}{2}-\frac{1}{2}\partial_{i}\phi\partial^{i}\phi
-\partial_{i}\varphi^{\dagger}\partial^{i}\varphi+M^2\varphi^{\dagger}\varphi+e^2\phi^2\varphi^{\dagger}\varphi
\right.
\nonumber\\
&+&\left.ef^{i}J_{i}-e^2f_{i}f^{i}\varphi^{\dagger}\varphi+\frac{m^2}{2}f_{i}f^{i}-\frac{m^2}{2}f_{0}f^{0}
-2m\phi{v_{i}f^{i}}\right),
\end{eqnarray}
and applying condition (\ref{vincsec}) to eliminate the $f^0$ operators, the Hamiltonian takes the form
\begin{equation}
H^{I}=H_{0}^{I}+H_{int\,1}^{I} + +H_{int\,2}^{I},
\end{equation}
being
\begin{eqnarray}
H_{0}^{I}&=&\int{d^2x}\left(\Pi^{I\dagger}\Pi^{I} + \frac{(\Pi_{\phi}^I)^2}{2}-\frac{1}{2}\partial_{i}\phi^{I}\partial^{i}\phi^{I} -2\phi^{I2} v_{0}^2
-\partial_{i}\varphi^{I\dagger}\partial^{i}\varphi^{I}+M^2\varphi^{I\dagger}\varphi^{I}\right.
\nonumber\\
&+&\left.\frac{m^2}{2}f_{i}^{I}f^{Ii}
-\frac{1}{2}\epsilon^{ij}\epsilon^{kl}(\partial_{i}f_{j}^{I})(\partial_{k}f_{l}^{I})
\right),
\end{eqnarray}
\begin{eqnarray}
  H_{int\,1}^{I}&=&-2\int{d^2x}\left(mv_{i}\phi^{I}{f^{Ii}}+v_{0}\phi^{I}\epsilon^{ij}\partial_{i}f_{j}^{I}\right),  
\end{eqnarray}
and 
\begin{eqnarray}
H_{int\,2}^{I}&=&\int{d^2x}\left(ef^{Ik}J_{k}^{I}-\frac{ie}{m}\epsilon^{kj}(\partial_{k}f_{j})(\Pi^{I}\varphi^{I}-\Pi^{I\dagger}\varphi^{I\dagger})-\frac{2ie}{m}v_{0}\phi^{I}(\Pi^{I}\varphi^{I}-\Pi^{I\dagger}\varphi^{I\dagger})
\right.
\nonumber\\
&+&\left.\frac{e^2}{2m^2}(\Pi^{I}\varphi^{I}-\Pi^{I\dagger}\varphi^{I\dagger})^2
-e^2f_{k}^{I}f^{Ik}\varphi^{I\dagger}\varphi^{I}+e^2\phi^{I2}\varphi^{I\dagger}\varphi^{I}\right).
\end{eqnarray}
In the Hamiltonian operator above, we write the superscript $I$ to indicate that the field operators belong to the interaction picture. 
Note that in the Hamiltonian $H^I_0$ the Lorentz symmetry breaking term generated a mass term for the scalar field $\phi$. 
In the Hamiltonian $H^{I}_{int\,1}$ we have the interaction of the scalar field $\phi$ with the gauge field $f$,  
whereas in the Hamiltonian $H^{I}_{int\,2}$ we have the interaction of the field $\varphi$ with the gauge field $f$ and with the scalar field $\phi$.
Then, the equations of motion that $\varphi^{I}$ and $\varphi^{I\dagger}$ obey are respectively given by:
\begin{eqnarray}
\dot{\varphi}(\vec{x})&=&i[H_{0}^{I},\varphi^{I}(\vec{x})]=\Pi^{I\dagger},
\\
\dot{\varphi}^{I\dagger}(\vec{x})&=&i[H_{0}^{I},\varphi^{I\dagger}(\vec{x})]=\Pi^{I}.
\end{eqnarray}
Thus, the propagator of the boson field in the momentum space is 
\begin{eqnarray}
\Delta(p)=\frac{i}{p^2 - M^2 +i\epsilon}.
\end{eqnarray}
For the self-dual field ($ f_i^I, i=1,2 $), the propagator is given by~\cite{Furtado:2008gs}
\begin{eqnarray}
\frac{1}{m^2}D^{lj}(k)=-\frac{i}{k^2-m^2+i\epsilon}\left(g^{lj}-\frac{k^{l}k^{j}}{m^2}
+\frac{i}{m}\epsilon^{lj}k_{0}-\frac{4v^{l}v^{j}}{k^2}-\frac{v^{l}k^{j}}{k^2}\right),
\end{eqnarray}
and for the mixed propagators, we have 
\begin{eqnarray}
\langle f^i\phi  \rangle=\frac{i}{k^2 - m^2 + i\epsilon}\left(\frac{2mv^i}{k^2}\right),
\end{eqnarray}
and
\begin{eqnarray}
\langle \phi\phi  \rangle=-\frac{i}{p^2 - m^2 +i\epsilon}\left( \frac{4m^2}{k^2} \right).
\end{eqnarray}

Therefore, we can now write the Hamiltonian of interactions in terms of the fundamental fields as follows:
\begin{eqnarray}
H_{int\,2}^{I}&=&\int{d^2x}\left(ief^{Ik}(\varphi^{I\dagger}\partial_{k}\varphi^{I}
-\partial_{k}\varphi^{I\dagger}\varphi^{I})-\frac{ie}{m}\epsilon^{lj}(\partial_{l}f_{j})
({\dot{\varphi}}^{I\dagger}\varphi^{I}-\varphi^{I\dagger}\dot{\varphi}^{I})
\right.
\nonumber\\
&+&\left.\frac{e^2}{2m^2}(\dot{\varphi}^{I\dagger}\varphi^{I}-\varphi^{I\dagger}\dot{\varphi}^{I})^2
-\frac{2ie}{m}v_{0}\phi^{I}(\dot{\varphi}^{I\dagger}\varphi^{I}-\varphi^{I\dagger}\dot{\varphi}^{I})\right.
\nonumber\\
&-&\left.e^2f_{k}^{I}f^{Ik}\varphi^{I\dagger}\varphi^{I}+e^2\phi^{I2}\varphi^{I\dagger}\varphi^{I}\right).
\end{eqnarray}
Note that the Hamiltonian of interactions above has six terms.
The first term corresponds to the spatial part of the dual field-current interaction. 
The second term, third term, and fourth term arise when we use the constraint condition to eliminate $f^0$ from the Hamiltonian.
Being the second term the interaction of the magnetic field with the time component of the current, while the third is the interaction of the temporal part of currents. 
Moreover, the extra terms are local in spacetime and not renormalizable by power counting.
The fourth term is the scalar field interacting with the temporal component of the current. 
The fifth term is the spatial part of the gauge-boson field interaction, and the sixth term is the scalar-boson field interaction. 

Here, we mention that the introduced Feynman rules are not manifestly covariant.
Therefore, we must clarify whether or not this can lead to a relativistically invariant $S$ matrix.
In the following, we will focus on testing the relativistic invariance of the model in connection with the specific process of elastic boson-boson scattering.
Since we are dealing with a non-renormalizable theory, we restrict our calculations to the tree approximation.

\section{boson-boson scattering}
\label{scat}
At this point, we will investigate the Lorentz invariance of the model. For this purpose, we will compute the $e^2$ order contribution to the lowest order elastic boson-boson scattering amplitude. 
Here, due to the non-renormalizability of the model, we will restrict the calculation to the tree-level approximation.
Hence, for the scattering amplitude, we have seven different types of terms grouped as follows:
\begin{eqnarray}
\label{eqs}
S^{(2)}=\sum_{\mu=1}^{7}{S^{(2)}_{\mu}},
\end{eqnarray}
where
\begin{eqnarray}
\label{amp1}
S^{(2)}_{1}&=&\frac{e^2}{2}\int\int{d^3xd^3y}\langle\Phi_{f}\mid T\{:[\varphi^{I\dagger}(x)\partial_{k}\varphi^{I}(x)
-\partial_{k}\varphi^{I\dagger}(x)\varphi^{I}(x)]f^{k}(x):
\nonumber\\
&\times &:[\varphi^{I\dagger}(y)\partial_{j}\varphi^{I}(y)
-\partial_{j}\varphi^{I\dagger}(y)\varphi^{I}(y)]f^{j}(y):\}\mid\Phi_{i}\rangle,
\end{eqnarray}
\begin{eqnarray}
\label{amp2}
S^{(2)}_{2}&=&-\frac{e^2}{m}\int\int{d^3xd^3y}\langle\Phi_{f}\mid T\{:[\varphi^{I\dagger}(x)\partial_{k}\varphi^{I}(x)
-\partial_{k}\varphi^{I\dagger}(x)\varphi^{I}(x)]f^{k}(x):
\nonumber\\
&\times &:[\epsilon^{lj}\partial_{l}f_{j}(y)
({\dot{\varphi}}^{I\dagger}(y)\varphi^{I}(y)-\varphi^{I\dagger}(y)\dot{\varphi}^{I}(y))]:\}\mid\Phi_{i}\rangle,
\end{eqnarray}
\begin{eqnarray}
\label{amp3}
S^{(2)}_{3}&=&\frac{e^2}{2m^2}\int\int{d^3xd^3y}\langle\Phi_{f}\mid T\{:[\epsilon^{kj}\partial_{k}f_{j}(x)
({\dot{\varphi}}^{I\dagger}(x)\varphi^{I}(x)-\varphi^{I\dagger}(x)\dot{\varphi}^{I}(x))]:
\nonumber\\
&\times &:[\epsilon^{il}\partial_{i}f_{l}(y)
({\dot{\varphi}}^{I\dagger}(y)\varphi^{I}(y)-\varphi^{I\dagger}(y)\dot{\varphi}^{I}(y))]:\}\mid\Phi_{i}\rangle,
\end{eqnarray}
\begin{eqnarray}
\label{amp4}
S^{(2)}_{4}&=&-\frac{ie^2}{2m^2}\int\int{d^3xd^3y}\delta(x-y)\langle\Phi_{f}\mid T\{:[{\dot{\varphi}}^{I\dagger}(x)\varphi^{I}(x)-\varphi^{I\dagger}(x)\dot{\varphi}^{I}(x)]:
\nonumber\\
&\times &:[{\dot{\varphi}}^{I\dagger}(y)\varphi^{I}(y)-\varphi^{I\dagger}(y)\dot{\varphi}^{I}(y)]:\}\mid\Phi_{i}\rangle,
\end{eqnarray}
\begin{eqnarray}
\label{amp5}
S^{(2)}_{5}&=&-\frac{2e^2v_{0}}{m}\int\int{d^3xd^3y}\langle\Phi_{f}\mid T\{:[\varphi^{I\dagger}(x)\partial_{k}\varphi^{I}(x)-\partial_{k}\varphi^{I\dagger}(x)\varphi^{I}(x)]f^{k}(x):
\nonumber\\
&\times &:[{\dot{\varphi}}^{I\dagger}(y)\varphi^{I}(y)-\varphi^{I\dagger}(y)\dot{\varphi}^{I}(y)]\phi(y):\}\mid\Phi_{i}\rangle,
\end{eqnarray}
\begin{eqnarray}
\label{amp6}
S^{(2)}_{6}&=&-\frac{2e^2v_{0}}{m^2}\int\int{d^3xd^3y}\langle\Phi_{f}\mid T\{:
[{\dot{\varphi}}^{I\dagger}(x)\varphi^{I}(x)-\varphi^{I\dagger}(x)\dot{\varphi}^{I}(x)]\phi(x):
\nonumber\\
&\times &:[{\dot{\varphi}}^{I\dagger}(y)\varphi^{I}(y)-\varphi^{I\dagger}(y)\dot{\varphi}^{I}(y)]
\epsilon^{kj}\partial_{k}f_{j}(y):\}\mid\Phi_{i}\rangle,
\end{eqnarray}
\begin{eqnarray}
\label{amp7}
S^{(2)}_{7}&=&\frac{e^2v_{0}^2}{m^2}\int\int{d^3xd^3y}\langle\Phi_{f}\mid T\{:
[{\dot{\varphi}}^{I\dagger}(x)\varphi^{I}(x)-\varphi^{I\dagger}(x)\dot{\varphi}^{I}(x)]\phi(x):
\nonumber\\
&\times &:[{\dot{\varphi}}^{I\dagger}(x)\varphi^{I}(y)-\varphi^{I\dagger}(y)\dot{\varphi}^{I}(y)]\phi(y): 
\}\mid\Phi_{i}\rangle.
\end{eqnarray}
Being $\mid\Phi_{i}\rangle$ and $\langle\Phi_{f}\mid$ the initial and final state of the process and $ T $ the chronological ordering operator. 
Here both $\mid\Phi_{i}\rangle$ and $\langle\Phi_{f}\mid$ are two-boson states.

In the next step, we compute partial amplitudes (\ref{amp1})-(\ref{amp7}) which in terms of initial $(p_{1},p_{2})$ and final $(p_{1}^{\prime},p_{2}^{\prime})$ moments become
\begin{eqnarray} 
\label{s1} 
S^{(2)}_{1}&=&-e^2N_{p}(2\pi)^3\delta(p_{1}^{\prime}+p_{2}^{\prime}-p_{1}-p_{2})\left[(p_{1}^{\prime}+p_{1})_{j}
(p_{2}^{\prime}+p_{2})_{l}\frac{1}{m^2}D^{jl}(k)+p_{1}\leftrightarrow p_{2} \right],
\\  
S^{(2)}_{2}&=&-e^2N_{p}(2\pi)^3\delta(p_{1}^{\prime}+p_{2}^{\prime}-p_{1}-p_{2})
\left[(p_{1}^{\prime}+p_{1})_{j}
(p_{2}^{\prime}+p_{2})_{0}\frac{1}{m^2}\Gamma^{j}(k)\right.
\nonumber\\
&+&\left.(p_{1}^{\prime}+p_{1})_{0}(p_{2}^{\prime}+p_{2})_{j}\frac{1}{m^2}\Gamma^{j}(-k)
+p_{1}\leftrightarrow p_{2} \right],
\\  
S^{(2)}_{3}&=&-e^2N_{p}(2\pi)^3\delta(p_{1}^{\prime}+p_{2}^{\prime}-p_{1}-p_{2})\left[(p_{1}^{\prime}+p_{1})_{0}
(p_{2}^{\prime}+p_{2})_{0}\frac{1}{m^2}\Lambda(k)+p_{1}\leftrightarrow p_{2} \right],
\\  
S^{(2)}_{4}&=&-e^2N_{p}(2\pi)^3\delta(p_{1}^{\prime}+p_{2}^{\prime}-p_{1}-p_{2})\left[(p_{1}^{\prime}+p_{1})_{0}
(p_{2}^{\prime}+p_{2})_{0}\frac{i}{m^2}+p_{1}\leftrightarrow p_{2} \right],
\\  
S^{(2)}_{5}&=&-e^2N_{p}(2\pi)^3\delta(p_{1}^{\prime}+p_{2}^{\prime}-p_{1}-p_{2})\left[(p_{1}^{\prime}+p_{1})_{j}
(p_{2}^{\prime}+p_{2})_{0}\left(\frac{v_{0}v^{j}}{m^4}\right)\right.
\nonumber\\
&+&\left.(p_{1}^{\prime}+p_{1})_{0}(p_{2}^{\prime}+p_{2})_{j}\left(\frac{v_{0}v^{j}}{m^4}\right)
+p_{1}\leftrightarrow p_{2} \right],
\\  
S^{(2)}_{6}&=&-e^2N_{p}(2\pi)^3\delta(p_{1}^{\prime}+p_{2}^{\prime}-p_{1}-p_{2})\left[(p_{1}^{\prime}+p_{1})_{0}
(p_{2}^{\prime}+p_{2})_{0}\left(\frac{k^{l}v_{l}}{m^4}\right)
+ p_{1}\leftrightarrow p_{2} \right],
\\  
S^{(2)}_{7}&=&-e^2N_{p}(2\pi)^3\delta(p_{1}^{\prime}+p_{2}^{\prime}-p_{1}-p_{2})\left[(p_{1}^{\prime}+p_{1})_{0}
(p_{2}^{\prime}+p_{2})_{0}\left(\frac{v_{0}^2}{m^2k^2}\right)
+p_{1}\leftrightarrow p_{2} \right],
\label{s7}
\end{eqnarray}
where we have
\begin{eqnarray}
&&\frac{1}{m^2}D^{jl}(k)=\frac{i}{k^2-m^2+i\epsilon}\left(-g^{jl}+\frac{k^{j}k^{l}}{m^2}-\frac{i}{m}
\epsilon^{jl}k_{0}-\frac{4v^jv^l}{k^2}-\frac{v^jk^l}{k^2}\right),
\\
&&\frac{1}{m^2}\Gamma^{j}(k)=\frac{i}{k^2-m^2+i\epsilon}\left(\frac{k^{j}k^{0}}{m^2}+\frac{i}{m}
\epsilon^{jl}k_{l}-\frac{4v^jv^0}{k^2}-\frac{v^0k^j}{k^2}\right),
\\
&&\frac{1}{m^2}\Lambda(k)=\frac{i}{k^2-m^2+i\epsilon}\left(-\frac{k^{l}k_{l}}{m^2}-\frac{4v^0v^0}{k^2}-\frac{v^0k^0}{k^2}\right),
\\
&&\frac{1}{m^2}\Sigma^j(k)=\frac{i}{k^2-m^2+i\epsilon}\left(-\frac{4v^0v^j}{k^2}  \right),
\\
&&\frac{1}{m^2}Q(k)=\frac{i}{k^2-m^2+i\epsilon}\left(-\frac{4v_l k^l}{k^2}  \right),
\\
&&\frac{1}{m^2}\Phi(k)=\frac{i}{k^2-m^2+i\epsilon}\left(-\frac{4v^0v^0}{k^2}  \right),
\end{eqnarray}
with
\begin{eqnarray}
k\equiv (p_{1}^{\prime}-p_{1})=-(p_{2}^{\prime}-p_{2}),
\end{eqnarray}
the momentum transfer. 
Now, by replacing equations (\ref{s1})-(\ref{s7}) into (\ref{eqs}) we get
\begin{eqnarray}
S^{(2)}=-e^2N_{p}(2\pi)^3\delta(p_{1}^{\prime}+p_{2}^{\prime}-p_{1}-p_{2})\left[(p_{1}^{\prime}+p_{1})_{\mu}
(p_{2}^{\prime}+p_{2})_{\nu}\frac{1}{m^2}D^{\mu\nu}(k)+p_{1}\leftrightarrow p_{2} \right],
\end{eqnarray}
being
\begin{eqnarray}
\frac{1}{m^2}D^{\mu\nu}(k)=-\frac{i}{k^2-m^2+i\epsilon}\left(g^{\mu\nu}-\frac{k^{\mu}k^{\nu}}{m^2}
+\frac{i}{m}\epsilon^{\mu\nu\alpha}k_{\alpha}-\frac{4v^{\mu}v^{\nu}}{k^2}-\frac{v^{\mu}k^{\nu}}{k^2}\right),
\end{eqnarray}
the propagator of the self-dual field $f^\mu$. 
Note that the scattering amplitude above is a Lorentz scalar and thus the model under consideration has passed the test of relativistic invariance at the quantum level.
Furthermore, in the tree-level approximation we can replace all non-covariant terms in the Hamiltonian of interactions by a minimum covariant interaction given by:
\begin{eqnarray}
H^I_{int\,2}=e J^I_\mu f^\mu, 
\end{eqnarray}
where
\begin{eqnarray}
J^I_\mu=i (\varphi^{I\ast}\partial_\mu\varphi^I - \partial_\mu\varphi^{I\ast}\varphi^I) - ef^I_\mu\varphi^{I\ast} \varphi^I - \frac{2m}{e}\phi v_\mu,
\end{eqnarray}
is the effective bosonic matter current. Moreover, by taking $v_{\mu}=0$ (without the effect of breaking the Lorentz symmetry), we have
\begin{eqnarray}
\frac{1}{m^2}D^{\mu\nu}(k)=-\frac{i}{k^2-m^2+i\epsilon}\left(g^{\mu\nu}-\frac{k^{\mu}k^{\nu}}{m^2}
+\frac{i}{m}\epsilon^{\mu\nu\alpha}k_{\alpha}\right),
\end{eqnarray}
with
\begin{eqnarray}
H^I_{int}=e J^I_\mu f^\mu, 
\end{eqnarray}
where
\begin{eqnarray}
J^I_\mu=i (\varphi^{I\ast}\partial_\mu\varphi^I - \partial_\mu\varphi^{I\ast}\varphi^I) - ef^I_\mu\varphi^{I\ast} \varphi^I.
\end{eqnarray}
Notice that in this way we recover the results found in~\cite{Anacleto:2000ea}.

\section{Dual Equivalence}
\label{Dual}
In this section, we study the duality between the self-dual and Maxwell-Chern-Simons models with Lorentz symmetry breaking coupled with bosonic matter. 
For this we write the Lagrangian as follows
\begin{eqnarray}
\label{acao2}
\Tilde{\cal L}&=&\frac{m}{2} \epsilon^{\mu\nu\rho}(f_{\mu}\partial_{\nu}f_{\rho})-\frac{\Tilde{m}^2}{2}f_{\mu}f^{\mu} 
+\frac{1}{2}\partial_{\mu}\phi\partial^{\mu}\phi+ 2m\phi v^{\mu}f_{\mu}
\nonumber\\
&+&\partial_{\mu}\varphi^{\ast}\partial^{\mu}\varphi-\mathcal{M}^2\varphi^{\ast}\varphi-ef_{\mu}J^{\mu},
\end{eqnarray}
where $\Tilde{m}^2=m^2-2e^2\varphi^{\ast}\varphi$ 
and $\mathcal{M}^2=M^2 + e^2\phi^2 $. Note that the mass parameters are field dependent, i.e., $\Tilde{m}$ depends on the charged scalar field and $\mathcal{M}$ depends on the real scalar field.

Equation of motion for the field $f_{\mu}$
\begin{eqnarray}
m\epsilon^{\mu\nu\rho}(\partial_{\nu}f_{\rho})
-\Tilde{m}^2f^{\mu}-e\tilde{J}^{\mu}=0,
\end{eqnarray}
where we have defined  $\Tilde{J}^{\mu}=J^{\mu} - 2e^{-1}m\phi v^{\mu}$.

Now, to determine the Euler vector we apply the following gauge transformation to the vector field $f_{\mu}$, such that 
$\delta f_{\mu}=\partial_{\mu}\epsilon$, being $\epsilon$ a parameter of gauge transformations. 
Thus, by varying the Lagrangian with respect to $f_{\mu}$, we find 
$\delta \mathcal{\tilde{L}}=K^{\mu}\partial_{\mu}$, where
\begin{eqnarray}
 K^{\mu}=m\epsilon^{\mu\nu\rho}(\partial_{\nu}f_{\rho})
-\Tilde{m}^2f^{\mu}-e\tilde{J}^{\mu}=0,   
\end{eqnarray}
is the Euler vector.

From the Euler-Lagrange equation, we obtain the equations for the fields $\phi$ and $\varphi$, given respectively by
\begin{eqnarray}  
\label{phi1}
\partial_{\mu}\partial^{\mu}\phi=2mv^{\mu}f_{\mu},
\end{eqnarray}
and
\begin{eqnarray}  
\label{varphi1}
(\partial_{\mu}\partial^{\mu}+\mathcal{M}^2)\varphi=
2ief^{\mu}\partial_{\mu}\varphi.
\end{eqnarray}
The next step is to establish duality between the self-dual and Maxwell-Chern-Simons models with Lorentz
symmetry breaking coupled with bosonic matter. Thus, we proceed by adopting the gauge embedding approach as done in~\cite{Anacleto:2001rp,Furtado:2008gs,Ferrari:2006vy}. In this way, we add to the original Lagrangian a term of the form $F(K^{\mu})$, as follows:
\begin{eqnarray}
\mathcal{\tilde{L}} \rightarrow \mathcal{\tilde{L}} + F(K^{\mu}),    
\end{eqnarray}
being $F(0)=0$. Now to obtain the form of the function $F(K^{\mu})$, we define
\begin{eqnarray}
 \mathcal{L}^{(1)}=\Tilde{\mathcal{L}} + B_{\mu}K^{\mu},   
\end{eqnarray}
where $B_{\mu}$ acts as a Lagrange multiplier. In addition, we assume that $\delta B_{\mu}=\partial_{\mu}\epsilon$ to cancel the variation of $\Tilde{\mathcal{L}}$. Therefore, proceeding as in Refs.~\cite{Anacleto:2001rp,Furtado:2008gs,Ferrari:2006vy} and after eliminating $B_{\mu}$, we find the following gauge invariant effective Lagrangian
\begin{eqnarray}
\mathcal{L}_{eff} =\Tilde{\mathcal{L}} + \frac{1}{2\Tilde{m}^2}K_{\mu}K^{\mu}.
\end{eqnarray}
Hence, renaming $f_{\mu}\rightarrow A_{\mu}$ to show the invariant character of the theory, we have
\begin{eqnarray}
\mathcal{L}_{eff}&=&\frac{1}{2} F_{\mu} F^{\mu}- \frac{m}{2}\epsilon^{\mu\nu\rho}A_{\mu}\partial_{\nu}A_{\rho} 
+\frac{1}{2}\partial_{\mu}\phi\partial^{\mu}\phi 
+\partial_{\mu}\varphi^{\ast}\partial^{\mu}\varphi-\mathcal{M}^2\varphi^{\ast}\varphi 
-\frac{em}{\Tilde{m}^2}\Tilde{J}_{\mu}F^{\mu} 
+ \frac{e^2}{2\Tilde{m}^2}\Tilde{J}_{\mu}\Tilde{J}^{\mu},
\end{eqnarray}
being,
\begin{eqnarray}
F^{\mu}\equiv \frac{1}{2}\epsilon^{\mu\nu\rho}F_{\nu\rho},    
\end{eqnarray}
the dual of the tensor $F_{\nu\rho}$.
By replacing $\Tilde{J}_{\mu}$ we obtain
\begin{eqnarray}
\mathcal{L}_{eff}&=&\frac{1}{2} F_{\mu} F^{\mu}- \frac{m}{2}\epsilon^{\mu\nu\rho}A_{\mu}\partial_{\nu}A_{\rho} 
+\frac{1}{2}\partial_{\mu}\phi\partial^{\mu}\phi +\frac{2m^2}{\Tilde{m}^2}\phi^2 v_{\mu}v^{\mu}
+\partial_{\mu}\varphi^{\ast}\partial^{\mu}\varphi-\mathcal{M}^2\varphi^{\ast}\varphi 
\nonumber\\
&-&\frac{em}{\Tilde{m}^2}{J}_{\mu} F^{\mu}
+\frac{2m^2\phi}{\Tilde{m}^2}{v}_{\mu}F^{\mu}
+\frac{e^2}{2\Tilde{m}^2}\left[{J}_{\mu}-\frac{4m}{e}\phi v_{\mu}\right]{J}^{\mu}.     
\end{eqnarray}
Note that a mass term for the field $\phi$ has been generated, which has also been previously obtained by the quantization procedure.

For this model, the equations for the $\phi$ field read
\begin{eqnarray}
\label{phi2}
\partial_{\mu}\partial^{\mu}\phi=2mv_{\mu} \left[\frac{m}{\Tilde{m}^2} F^{\mu}+\frac{2m}{\Tilde{m}^2}\phi v^{\mu}-\frac{e}{\Tilde{m}^2}J^{\mu}\right] ,
\end{eqnarray}
and for the $\varphi$ field, we have
\begin{eqnarray}
\label{varphi2}
 \left(\partial_{\mu}\partial^{\mu}+\mathcal{M}^2\right)\varphi= -2ie\left[\frac{m}{\Tilde{m}^2}F^{\mu} 
 +\frac{2m}{\Tilde{m}^2}\phi{v}^{\mu}
 -\frac{e}{\Tilde{m}^2}{J}^{\mu} 
 \right] \partial_{\mu}\varphi.
\end{eqnarray}
Finally, by comparing equations \eqref{phi1}, \eqref{varphi1}, \eqref{phi2} and \eqref{varphi2}, we obtain that the correct map 
between the self-dual model and the Maxwell-Chern-Simons is given by
\begin{eqnarray}
 f^{\mu}\rightarrow\frac{m^2}{\Tilde{m}^2}\left(\frac{F^{\mu}}{{m}} +\frac{2\phi}{{m}}{v}^{\mu}
 -\frac{e}{{m}^2}{J}^{\mu}\right).
\end{eqnarray}
In this way we have successfully completed the dual mapping between the two models. The mapping obtained here is a generalization of the results obtained in~\cite{Anacleto:2001rp} and~\cite{Furtado:2008gs}. 
By setting $v^{\mu}=0$ (in the absence of Lorentz symmetry breaking) we will recover the results obtained in~\cite{Anacleto:2001rp}. 
\begin{eqnarray}
 f^{\mu}\rightarrow\frac{m^2}{\Tilde{m}^2}\left(\frac{F^{\mu}}{{m}}-\frac{e}{{m}^2}{J}^{\mu}\right).
\end{eqnarray}    
On the other hand, without the presence of the field $\varphi$, but maintaining the effect of breaking the Lorentz symmetry, we have the result found in~\cite{Furtado:2008gs}.
\begin{eqnarray}
 f^{\mu}\rightarrow\left(\frac{F^{\mu}}{{m}} +\frac{2\phi}{{m}}{v}^{\mu}\right).
\end{eqnarray}
{Finally, in our analysis, we have generalized the result that was investigated in~\cite{Anacleto:2000ea} by including the contribution of the breaking of Lorentz symmetry to the quantization and scattering process. We have also generalized the dual mapping in~\cite{Furtado:2008gs} by considering the coupling with the bosonic field. 
The results obtained here may be useful in condensed matter, such as the quantum Hall effect, high-temperature superconductivity, and vortex solutions among others.}

\section{Summary}
\label{summ}
In this work, we have considered the dynamics, at the quantum level, of the self-dual model coupled to Lorentz symmetry-breaking bosons. Then, after eliminating the degree of freedom $f^0$ the equal-time commutators with the bosonic fields that are not zero, we have formulated the dynamics of the interaction framework for the self-dual model minimally coupled to the bosons by breaking the Lorentz symmetry. As a consequence, 
we have shown that a mass term is generated for the real scalar field $\phi$.
However, we prove that the non-covariant parts in $H^I_{int\, 2}$ are equivalent to the minimum covariant field-current interaction. The high-energy behavior of the $f$ propagator signals that the coupled theory is not renormalizable. 
In this way, we have shown that relativistic invariance of the model is preserved at the quantum level.
We finish the paper by establishing dual equivalence between the self-dual and Maxwell-Chern-Simons models with Lorentz symmetry breaking and coupled with bosonic matter field.

\acknowledgments

We would like to thank CNPq, CAPES and CNPq/PRONEX/FAPESQ-PB (Grant nos. 165/2018 and 015/2019),
for partial financial support. M.A.A., F.A.B. and E.P. acknowledge support from CNPq (Grant nos. 306398/2021-4, 309092/2022-1, 304290/2020-3).
\\
\\
\textbf{Data Availability Statement:} No Data associated in the manuscript.

\end{document}